\documentclass{elsart}

\usepackage{graphicx}
\usepackage[latin1]{inputenc}
\usepackage{natbib,amssymb,graphicx}
\usepackage{pstricks} 

\usepackage[dvips]{hyperref}

\newcommand{\pT}{p_T}

\newcommand{\Pythia}{\textsc{Pythia}}
\newcommand{\Jetset}{\textsc{Jetset}}
\newcommand{\Fritiof}{\textsc{Fritiof}}

\newcommand{\UNIT}[1]{\mbox{$\,{\rm #1}$}}

\newcommand{\MeV}{\UNIT{MeV}}
\newcommand{\MeVc}{\UNIT{MeV/c}}
\newcommand{\GeV}{\UNIT{GeV}}
\newcommand{\GeVc}{\UNIT{GeV/c}}

\newcommand{\REM}[1]{}

\begin{document}

\begin{frontmatter}

\title{Production of charged pions off nuclei with 3$\cdots$30 GeV
  incident protons and pions\thanksref{label1}}
\thanks[label1]{Work supported by DFG.}

\author{K.~Gallmeister}
\ead{Kai.Gallmeister@theo.physik.uni-giessen.de}
\author{U.~Mosel}
\address{Institut f\"ur Theoretische Physik, Universit\"at
  Giessen, Germany}

\begin{abstract}
  We compare calculations for the production of charged pions by pion
  or proton beams off nuclei calculated within our coupled channel
  transport model (GiBUU) with recent data of the HARP collaboration
  for beam energies from 3 up to 13\GeV{}. Predictions for the 30
  \GeV{} data for pions and kaons from the NA61/SHINE experiment are
  included.
\end{abstract}

\begin{keyword}

hadron formation \sep hadron induced high-energy interactions \sep
  meson production

\PACS 13.75.-n \sep 13.85.-t \sep 25.40.-h \sep 25.80.-e

\end{keyword}

\end{frontmatter}

\section{Introduction}

Recently the HARP experiment has published data for $\pi^\pm$
production by protons or pion beams in the momentum range
3$\cdots$13\GeVc{} impinging on different nuclear targets
\cite{Catanesi:2008zz,Catanesi:2008zzb,Catanesi:2005rc}.  Here the
main goal is to contribute to the understanding of the neutrino fluxes
of accelerator neutrino experiments such as K2K, MiniBooNE and
SciBooNE or for a precise calculation of the atmospheric neutrino
fluxes. Some of the experimental data were also compared to several
different generator models used in GEANT4 and MARS simulation packages
\cite{Catanesi:2008zzb}.  The overall agreement was reasonable, while
for some models discrepancies up to factors of three were seen.
Unfortunately, none of these models is applicable for all energies
considered in the experiment: somewhere at 5$\cdots$10\GeV{} a
distinction between low energies and high energies has to be done,
limiting the range of validity of these models. The lack of good
quality and systematic data concerning hadron--nucleus collisions in
this energy regime has for long been an obstacle for a serious test of
the models. The advent of the HARP experiment has changed the
situation, since it offers charged pion double differential cross
sections with a good systematics in angle, pion momentum, incident
energy and target mass.

In this paper we show comparisons of calculations within the GiBUU
transport model to data obtained
with the forward spectrometer as well as with the large-angle
spectrometer of the HARP experiment. The calculations are done without
any fine tuning to the data covered here with the default parameters
as used in the GiBUU framework for all possible processes.

\section{GiBUU Transport Model}
%

In this paper we employ the GiBUU model for an analysis of these data.
This model has been developed as a transport model for energies from
some \MeV{} up to tens of \GeV{} \cite{GiBUU}. Here we can study all
kind of elementary collisions induced by baryons, mesons (see
e.g.~\cite{Buss:2006yk}), (real and virtual) photons (see
e.g.~\cite{Buss:2006vh,Falter:2004uc}) and neutrinos (see
e.g.~\cite{Leitner:2006ww,Leitner:2008fg} and further references
therein) on all kind of nuclei within a unified framework.
The underlying code is written in modular FORTRAN 2003 and available
for download at \cite{GiBUU}.

In the GiBUU model we solve a strongly coupled system of equations for
one--particle phasespace densities, the so called
Boltzmann--Uehling--Uhlenbeck (BUU) equations. The total time
evolution of these phasespace densities is given by the motion in
hadronic mean fields and potentials (Vlasov--part) combined with a
collision integral. The actual implementation includes 61 baryons and
21 mesons. The evolution equations for all these particles are coupled
by the Hamiltonian in the Vlasov--part and also by the collision
integral.
The phase space densities are approximated via the testparticle ansatz
as a sum of delta distributions in spatial dimensions and momentum.

The collision integral is mainly dictated by (elastic and inelastic)
two body collisions. At low energies these reaction mechanisms are
governed by a resonance description. At higher energies, the
collisions are done via the \Pythia{} event generator. Here we are
switching smoothly between the two descriptions for
$\sqrt{s}=2.2(2.6)\pm0.2\GeV$ for meson-baryon (baryon-baryon)
collisions.  Contrary to most other hadronic interaction models, which
are using a string excitation picture via Pomeron exchange
(e.g.~\Fritiof{} or some own implementations), we are using \Pythia{}
(v6.4) also at these low energies.  At the energies covered here, we
are in a transition region, where the high energy processes embedded
in the \Pythia{} model may reach their limits. However, \Pythia{} also
contains some low energy processes. Contrary to a Pomeron exchange
model, as e.g.~implemented in \Fritiof{} or UrQMD, low energy
interactions are treated as string flips in \Pythia{}.  Thus only the
\Jetset{}--Part of \Pythia{} is responsible for these processes and,
therefore, the influence and number of (internal) model parameters is
limited.  We have ascertained that \Pythia{} describes the hadron
production cross sections in general quite well even at relatively low
invariant masses; the detailed comparison with the HARP data can serve
as a further test of the general accuracy of the method.  For further
details, including e.g.~plots of total cross sections for the
elementary processes see Ref.~\cite{GiBUU}.

The elementary interaction with the nucleon is assumed to be the same
as that with a free nucleon. All the standard nuclear effects like
Fermi motion, Pauli blocking and nuclear shadowing are properly taken
into account. In a second step, all produced (pre--)hadrons are
propagated through the nuclear medium according to the semi-classical
Boltzmann--Uehling--Uhlenbeck transport equation. The concept of
pre-hadrons (i.e.\ produced hadrons interact with some reduced cross
section during their hadronization time) was introduced in order to
realize color transparency and formation time effects
\cite{Gallmeister:2005ad,Gallmeister:2007an}. All interactions,
primary or secondary, are therefore treated within the same
prescription and we thus present a full consistent coupled channel
transport approach.

At the given energies (beam momenta $>3\GeVc{}$), these elementary $pN$
or $\pi^\pm N$ interaction are done within the \Pythia{} prescription,
since the available collision energies are above the resonance regions
($\sqrt{s}_{\pi N,pN}=2.8\cdots4.9\GeV{}$). In addition also the
momenta of the observed final particles are quite large (larger than
hundreds of \MeVc{}), thus reducing the importance of the baryonic
potentials, which are a major difference of the GiBUU model to other
cascade Monte Carlo models.

\section{Results}
%

\begin{figure}[tb]
  \begin{center}
    \hspace*{\fill}%
    \includegraphics[width=6.5cm,clip=true]{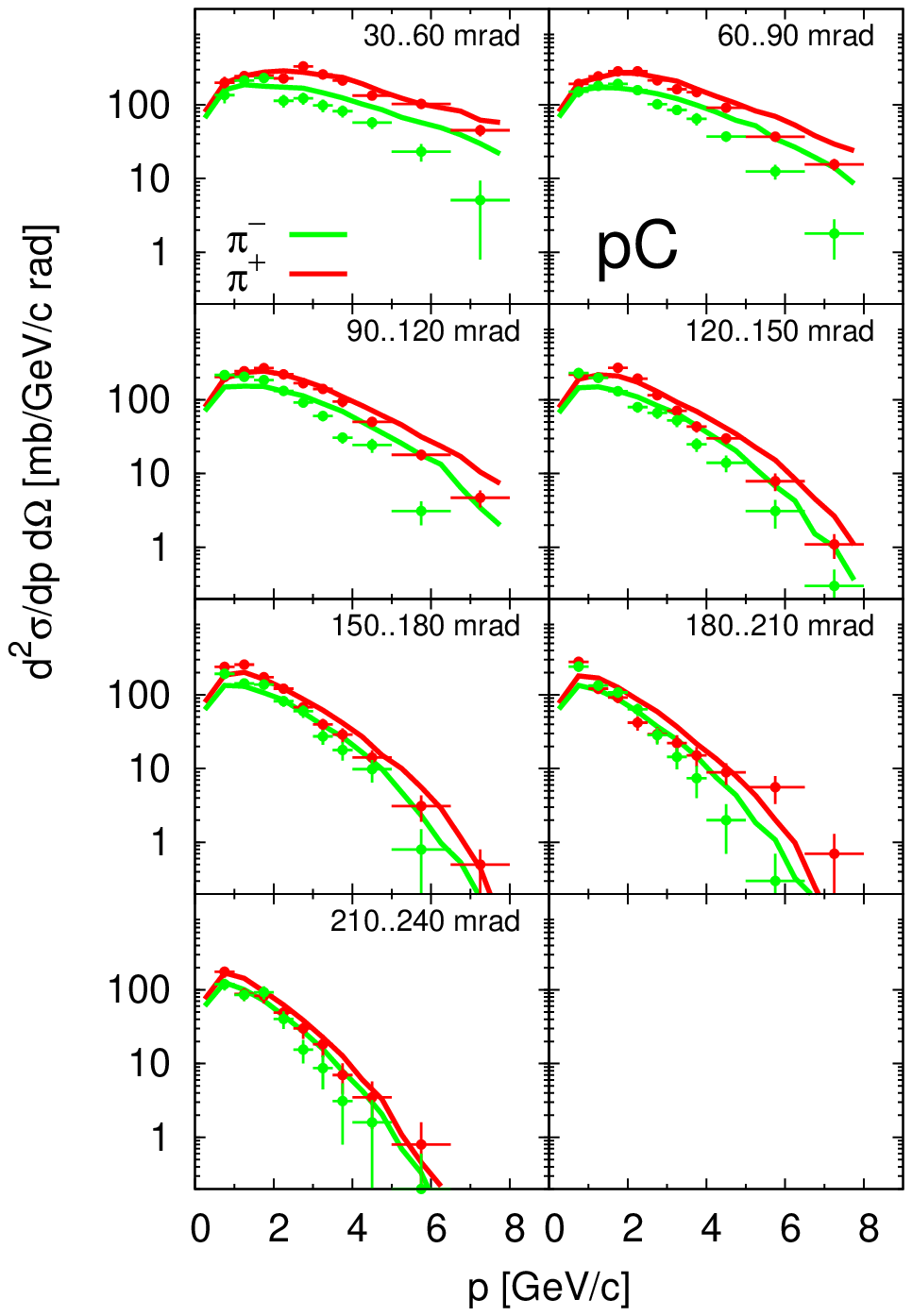}
    \hspace*{\fill}%
    \includegraphics[width=6.5cm,clip=true]{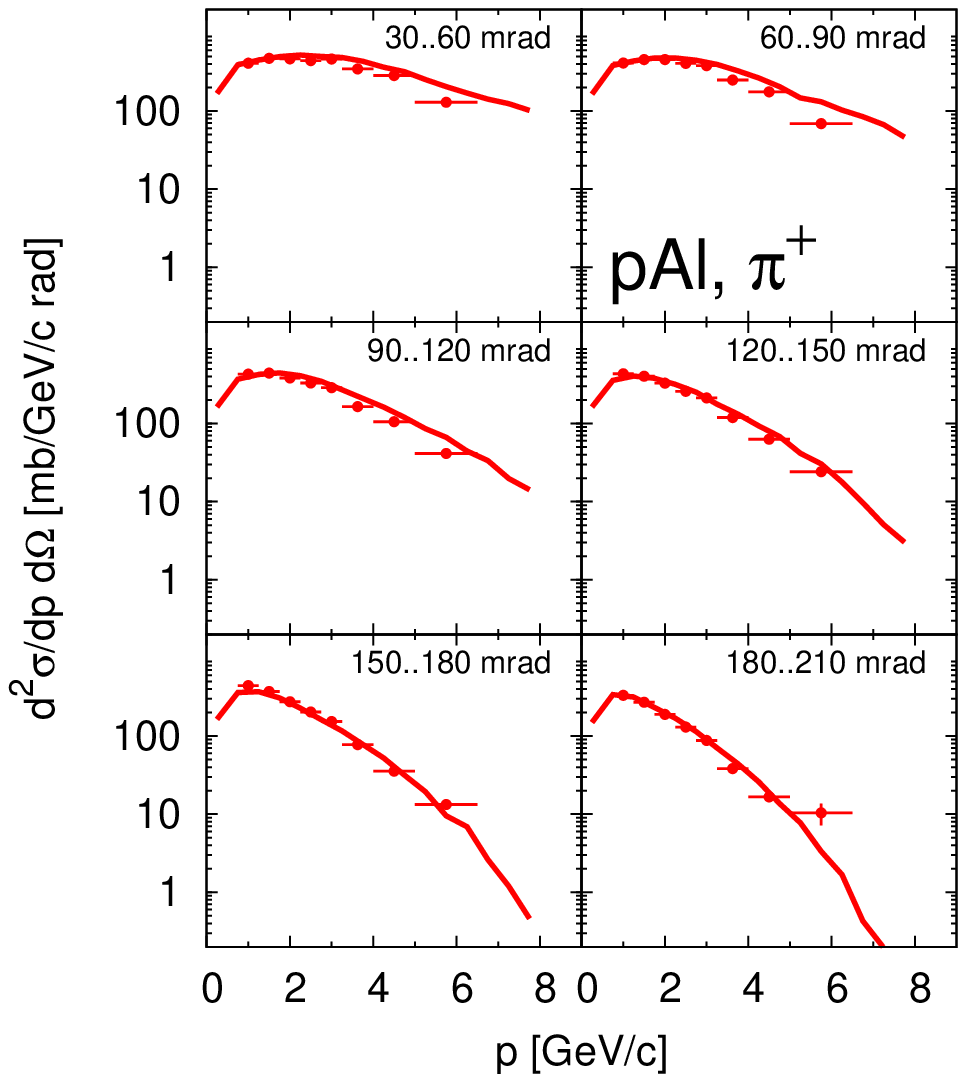}
    \hspace*{\fill}%

    \caption%
    {\textit{Cross section $d^2\sigma/dp\,d\Omega$ for
        $p+C\to\pi^\pm+X$ with 12\GeVc{} beam momentum (left) and
        $p+Al\to\pi^++X$ with 12.9\GeVc{} beam momentum (right).
        Experimental data are from
        \cite{Catanesi:2008zz,Catanesi:2005rc} (HARP small angle
      analysis).}}
  \label{fig:Harp_HiC_p12}
\end{center}
\end{figure}

We begin our comparison with data obtained with the forward
spectrometer for a Carbon target \cite{Catanesi:2008zz}. In
fig.~\ref{fig:Harp_HiC_p12} we show the double differential production
cross section of charged pions for the 12\GeVc{} proton beam.  Here the
covered momentum range of the detected pions reaches up to 8\GeVc{}.
For more backward angles ($\gtrsim 120\UNIT{mrad}$) the agreement is
perfect. For the very forward angles (up to 120\UNIT{mrad}) we observe
a too flat behavior of the calculated spectra compared to data.
The calculations overshoot the data for large momenta significantly,
especially for the $\pi^-$ channel\footnote{During the final prepation
  of this paper, experimental data of the HARP collaboration on
  $^{14}$N and $^{16}$O was released, showing a strong dependence of
  the very forward data point on the nuclear size, with a much smaller
  disagreement with our calculations for the $^{16}$O target.}. This
seems to be a general shortcoming of all the hadronic interaction
models GHEISHA, UrQMD and DPMJET-III shown in \cite{Catanesi:2008zz}.
However, in contrast to the model predictions of these models, our
calculations reproduce the decrease of the cross section for very
small momenta (e.g.\ for $p\lesssim2\GeVc$ for angles smaller than
90\UNIT{mrad}).

In fig.~\ref{fig:Harp_HiC_p12} (right panel) we also show a comparison
of our model calculations with experimental data
\cite{Catanesi:2005rc} with a 12.9\GeVc{} beam on an Al target for
positive charged pions. Here the agreement is very good for all
momenta and angles.

\begin{figure}[tb]
  \begin{center}
    \hspace*{\fill}%
    \includegraphics[height=6.0cm,clip=true]{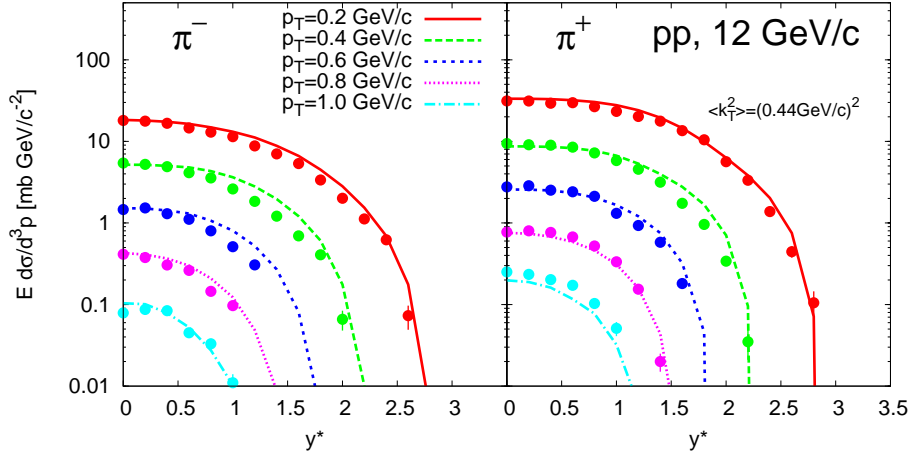}
    \hspace*{\fill}%

    \caption%
    {\textit{Invariant cross section $E\, d\sigma/d^3p$
      for $pp\to\pi^\pm X$ with 12\GeVc{} beam
      momentum. Experimental data are from
      \cite{Blobel:1973jc}. Due to symmetry only $y^*>0$ is shown.}}
  \label{fig:Blobel12}
\end{center}
\end{figure}

In order to clarify the question whether the problems observed for the
very forward angles and high momenta for $^{12}$C are due to problems
in the treatment of the FSI or already present in the hard first
interaction, we compare in fig.~\ref{fig:Blobel12}
our model with experimental data for elementary reactions
\cite{Blobel:1973jc}.
Here the relevant kinematical variables are the transverse momentum
$\pT$ and the cm rapidity $y^*$. It is obvious that the actual model
implementation agrees very well with these data, even though there are
small discrepancies visible around $y^*\sim 1.5$.  It is worthwhile to
mention, that this holds only for the newest versions of \Pythia{}
(v6.4) as used here in our model; older versions (e.g.~v6.2) seem to
prefer low transverse momentum and high rapidity and are not able to
describe these data. Since we are able to describe the outgoing pions
from elementary $pN$ collision correctly, the overprediction of the
cross section at high momenta and very forward angles on nuclear
targets has to be attributed to interactions of the outgoing pion with
the the nuclear medium.  However, as we will see below, our
calculations are in agreement with pion induced data on nuclei so that
a puzzle remains. We leave the further investigation of this problem
to future studies \cite{Gallmeister:inProgress}.

\begin{figure}[tb]
  \begin{center}
    \hspace*{\fill}%
    \includegraphics[height=7.0cm,clip=true]{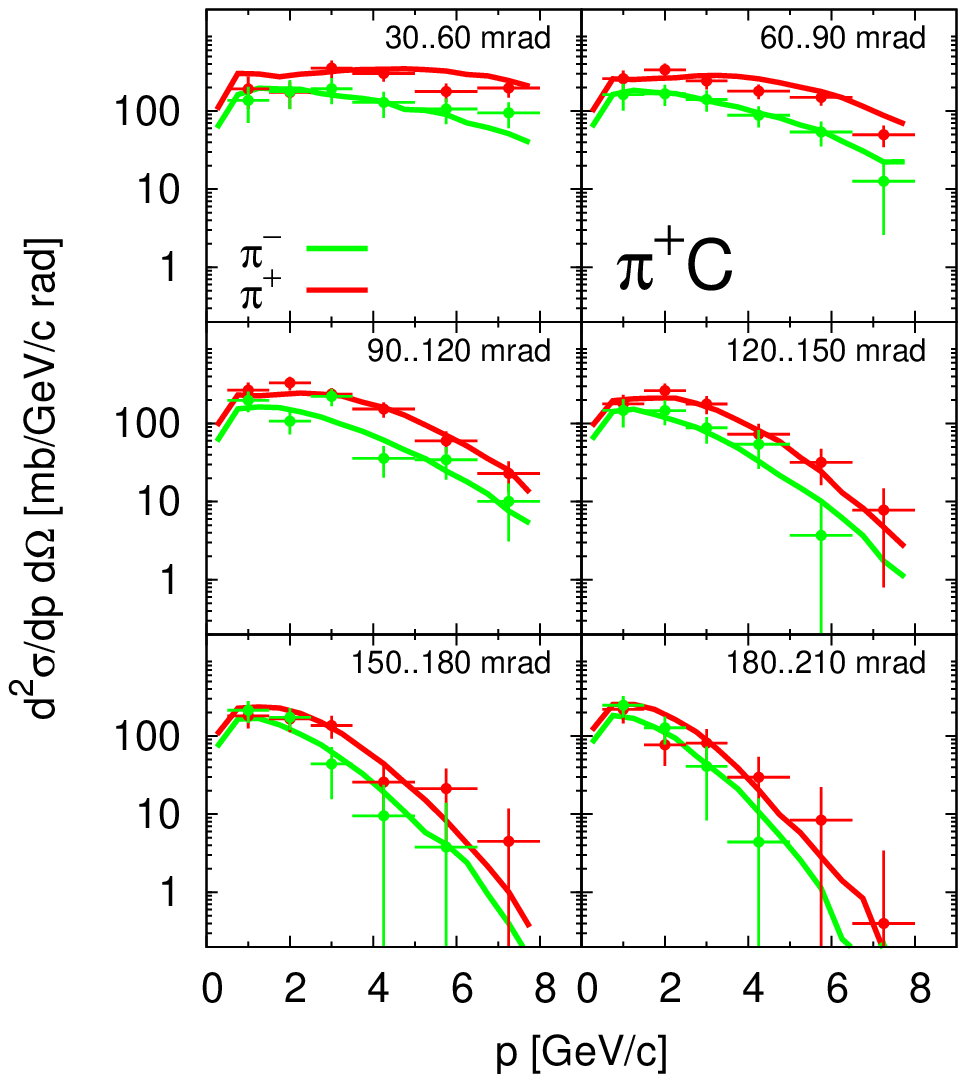}
    \hspace*{\fill}%
    \includegraphics[height=7.0cm,clip=true]{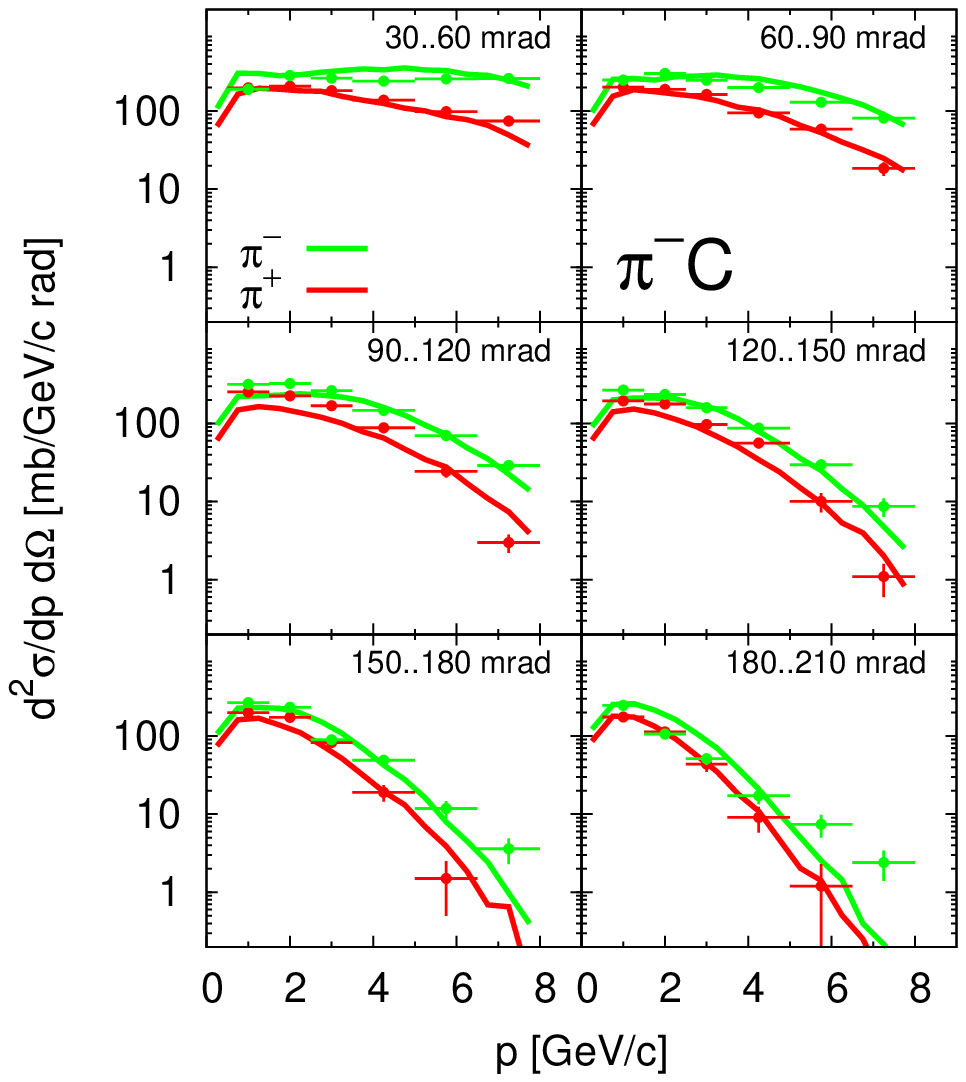}
    \hspace*{\fill}%

    \caption%
    {\textit{Cross section $d^2\sigma/dp\,d\Omega$
      for $\pi^\pm+C\to\pi^\pm+X$ with 12\GeVc{} beam
      momentum. Experimental data are from
      \cite{Catanesi:2008zz} (HARP small angle analysis).}}
  \label{fig:Harp_HiC_pi12}
\end{center}
\end{figure}

In fig.~\ref{fig:Harp_HiC_pi12} we show the comparison with data with
$\pi^+$ and $\pi^-$ beams.
Here the agreement is very good for all available angles and all
momenta.  Not only the charge--conserving channels, but also the
double charge exchange channels are well reproduced. This limits
the uncertainties connected with the final state interactions in the
proton induced reactions, as discussed before.

We continue our comparison with data with the large angle spectrometer
\cite{Catanesi:2008zzb}. In order to keep this paper reasonably short
we restrict ourselves to comparisons for a few selected energies only.
A gallery of more comparisons is available at \cite{GiBUUGallery}.

\begin{figure}[tb]
  \begin{center}
    \hspace*{\fill}%
    \includegraphics[height=12cm,clip=true]{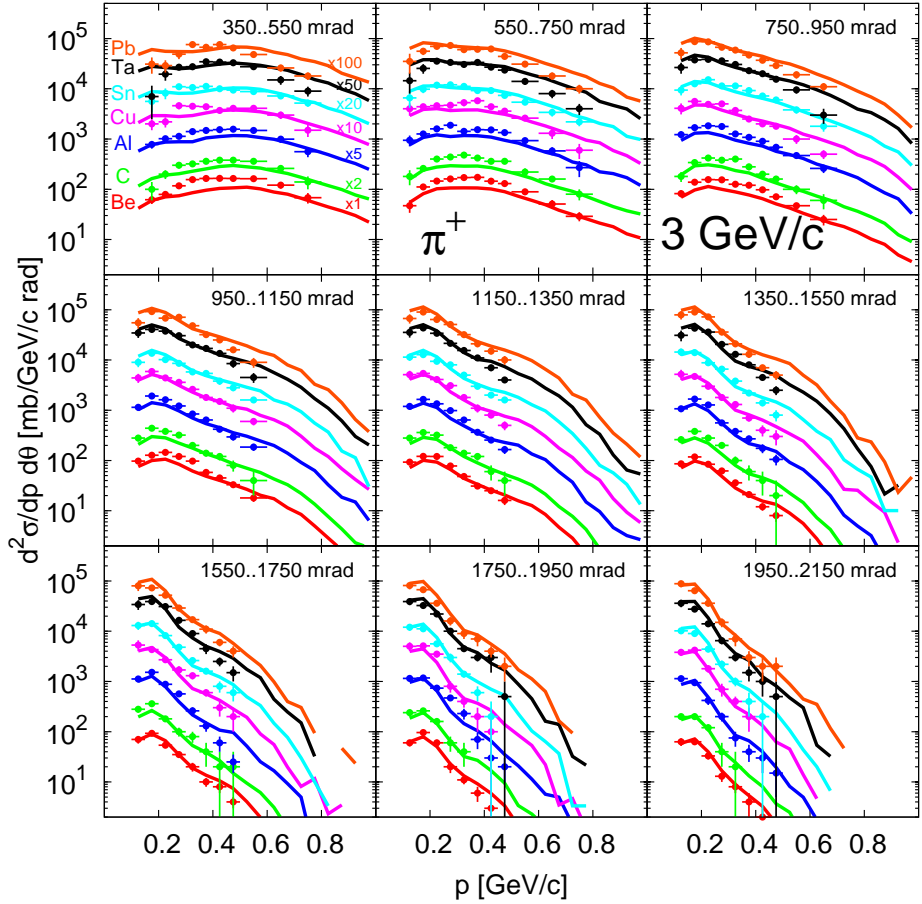}
    \hspace*{\fill}%

    \caption%
    {\textit{Cross section $d^2\sigma/dp\,d\theta$ for $p+A\to\pi^++X$
        with 3\GeVc{} beam momentum. Experimental data are from
        \cite{Catanesi:2008zzb} (HARP large angle analysis), curves
        and data scaled as indicated.}}
    \label{fig:08_31_Harp_03}
  \end{center}
\end{figure}

In fig.~\ref{fig:08_31_Harp_03} we compare calculations with the data
for the proton beam at 3\GeV{}. In the large angle analysis all the
momenta of the detected pions are below 1\GeVc{}.
One sees a very good overall agreement for perpendicular or even
backward directions for all nuclei. Small discrepancies occur mainly
for angles below 750\UNIT{mrad} at very low momenta $\lesssim 0.2
\GeVc$ where the calculations are higher than the experimental data.
Correspondingly, the slope for momenta larger than 0.4\GeVc{} is too
flat in our calculations. For light nuclei the slope is in agreement
with data, while the overall yield is somewhat too small. We note that
these observations also hold for the negatively charged pions not
shown here.

\begin{figure}[tb]
  \begin{center}

    \hspace*{\fill}%
    \includegraphics[height=12cm,clip=true]{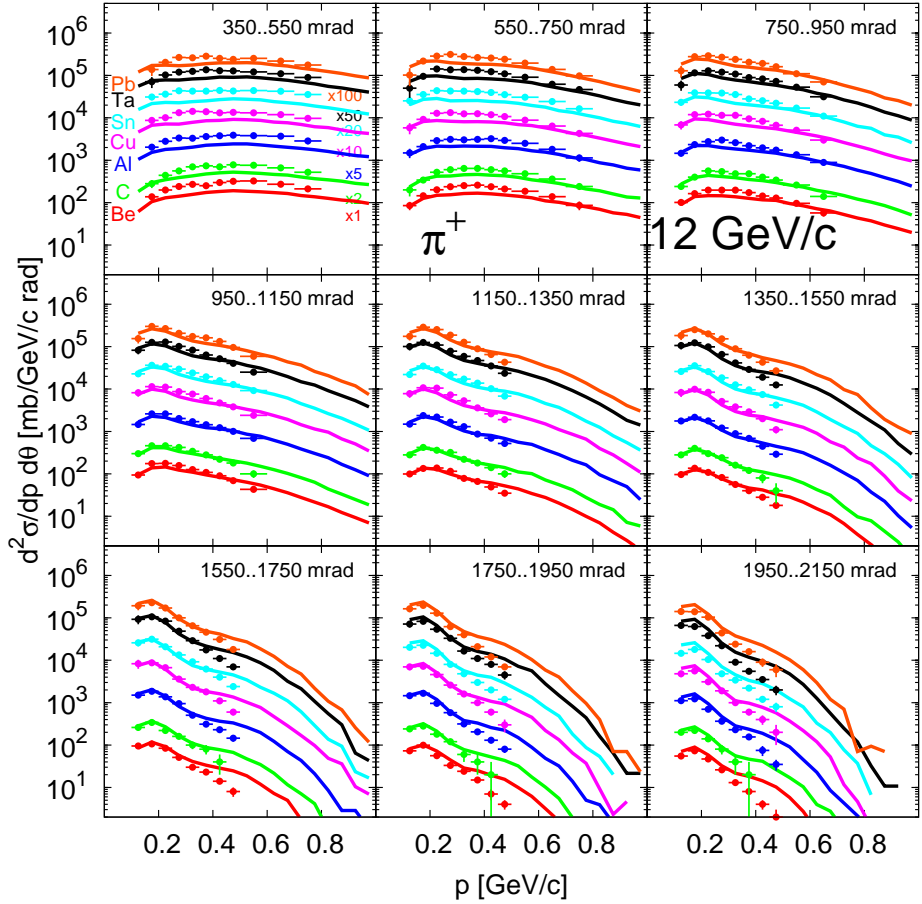}
    \hspace*{\fill}%

    \caption%
    {\textit{Cross section $d^2\sigma/dp\,d\theta$ for $p+A\to\pi^++X$
        with 12\GeVc{} beam momentum. Experimental data are from
        \cite{Catanesi:2008zzb} (HARP large angle analysis), curves
        and data are scaled as indicated.
        The targets are indicated in the top-left frame.}}
    \label{fig:08_31_Harp_12}
  \end{center}
\end{figure}

In order to illustrate the energy dependence of our results, we
compare in fig.~\ref{fig:08_31_Harp_12} the calculations for positive
pion production with the 12\GeVc{} proton beam.
The overall behavior of the calculations changes smoothly from
3\GeVc{} to 12\GeVc{}, a comparison for 5 and 8\GeVc{} can be found in
\cite{GiBUUGallery}.  For the higher energies the data do not show the
strong dip observed for small angles and small momenta at 3\GeVc{}.
However the overall yield for the small angles is still somewhat too
low.

For all energies one observes for the perpendicular directions
($\simeq 1550\UNIT{mrad}$) a 'bumpy' structure around $p \approx
0.5\GeVc$.  We note, that while this structure is not very pronounced
in the experimental data for $\pi^+$, the experimental data for the
$\pi^-$ channel (not shown here) do exhibit this feature. Calculations
for a nucleon target indicate a smooth behavior. For the nuclear
target at momenta around $0.2\GeVc$ rescattering and the $\Delta$
resonance dominate. This small momentum regime is populated by
originally higher-energy pions that have been slowed down due to
rescattering; only due to these final state interactions the overall
yield at the lower momenta is reproduced. Without FSI the yield for
momenta around $0.2\GeV$ is underestimated by at least one order of
magnitude.

We note, that the laboratory angles of the HARP experiment can be
translated into a cm pseudo rapidity $\eta^*$ \footnote{We keep this
  label different from the 'cm rapidity' $y^*$ as used in connection
  with the data from ref.~\cite{Blobel:1973jc}, since they differ at
  smaller momenta due to the inclusion of the particle mass. In
  adition we indicate by the '*', that we are discussing these values
  in the $NN$ center of momentum frame.}.  The given angular bins of
the forward angular spectrometer analysis cover the range
$\eta^*=0.5\cdots2.1$. The three smallest angles correspond to the
range $\eta^*>1.2$.  The discrepancies for large angles of the
backward spectrometer analysis then show up as discrepancies at
rapidities $\eta^*<-1.8$. This is the feature already seen in the
comparisons to the small angle analysis, but now for the very backward
rapidities.  The transverse momentum region in the data for these
extreme rapidities is comparable.

In order to quantify the agreement between data and calculations, we
calculated $\chi^2/n_{\rm dof}$ values for every curve shown in this
paper. We note, however, that our
model does not exhibit any free parameter and we did not any fit
procedure to minimize this measure. The resulting values for
$\chi^2/n_{\rm dof}$ may thus be very large. Nevertheless, they show
some remarkable systematics, as e.g.~they reach up for all nuclei to
$\sim 20$ for $\theta<950\UNIT{mrad}$ and also for
$\theta>1550\UNIT{mrad}$ with proton beam and the large angle analysis
(cf.~fig.~\ref{fig:08_31_Harp_03} and fig.~\ref{fig:08_31_Harp_12}).
Inbetween the measure drops to values around 4. This is reflected in
the fact, that for the small angles of this backward analysis all
calculated curves show the right slope, but are systematically
below the data for all nuclei. In fact, by introducing an additional
artifical overall normalization factor, we would be able to decrease
the $\chi^2/n_{\rm dof}$ values down to 1.  The mismatch for the
largest angles could not be cured by simple up-/downscaling the
calculated curves. Here the large numerical values represent the fact,
that we have significant discrepancies in the slopes, as discussed
above. The measure for the forward angle analysis does not show any
simple tendencies, except the fact that the pion induced data yield
values $\sim3$ in the negative channel and $\sim1$ in the positive one and
are thus in a reasonable regime.

\begin{figure}[tb]
  \begin{center}

    \hspace*{\fill}%
    \includegraphics[height=8cm,clip=true]{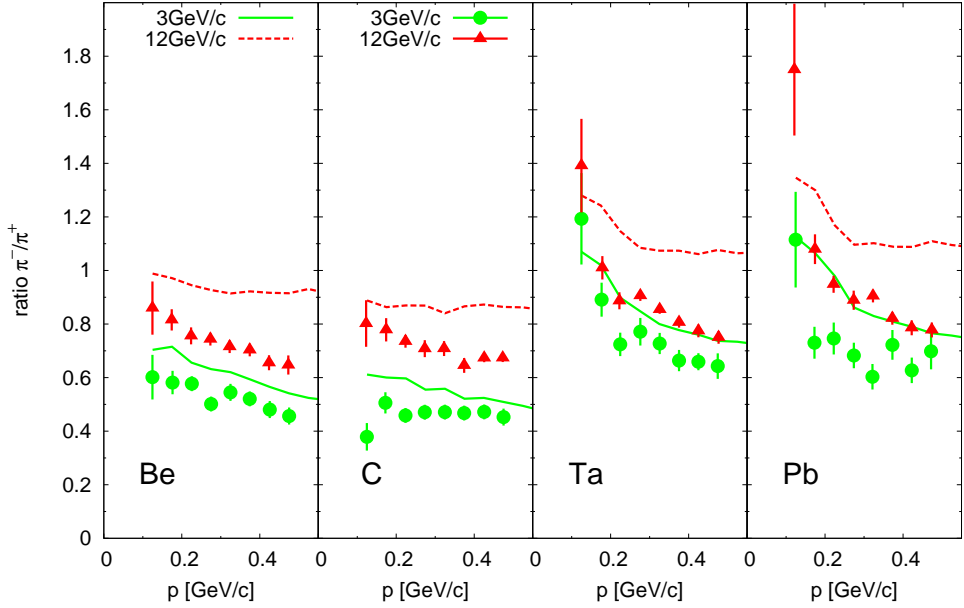}
    \hspace*{\fill}%

    \caption%
    {\textit{Ratio of cross section for $\pi^-$ over $\pi^+$ for
        p+Be, p+C, p+Ta and p+Pb interactions at 3 and 12\GeVc{} as
        function of the momenta of the pions, integrated over the
        angles 350\UNIT{mrad} to 1550\UNIT{mrad}. Data are from
        \cite{Catanesi:2008zzb}.}}
    \label{fig:08_31_Harp_ratio}
  \end{center}
\end{figure}

While the experimental spectra of positive and negative charged pions
show substantial differences, the spectra resulting from our
calculations do exhibit the same features for both charge states.  In
order to quantify the differences between the two isospin states, we
show in fig.~\ref{fig:08_31_Harp_ratio} the ratio of the yields of
$\pi^-$ over the yield of $\pi^+$.
We have chosen the two smallest and the two largest nuclei. The
discrepancies between data and calculations are quite significant,
although the overall tendencies are reproduced. As in experiment we
observe an excess of $\pi^-$ over $\pi^+$ for large nuclei at low
momenta. It was suggested by the experiment E910
\cite{Chemakin:2001qx} that this is due to production of $\Lambda^0$.
However, this is not supported by our calculations where the charge
asymmetry for the pions originates in decays of $\Delta$'s at rest.
The latter have long collision histories: The higher the initial
energies, the more collisions it takes to generate resonances
\textit{at rest}, i.e.~the higher the initial energy, the longer the
collsion history in order to 'stop' the resonance. If there is an
imbalance of the charges of the collision partners (more neutrons than
protons), the charge of the outgoing particles is driven towards the
charge asymmetry. Therefore we observe that the pion charge asymmetry
grows with 1) the initial energy (more collisions needed to get a
stopped Delta) and with 2) the neutron/proton asymmetry.  We have
checked, that in a (fictitious) charge symmetric nucleus of the size
of Pb the ratio of negative over positive charged pions is indeed the
same as in Be. We have also checked, that our results are not affected
by switching on or off baryonic potentials or Coulomb corrections.

At the same time the HARP experiment is used to reduce uncertainties
on the flux calculations of the K2K experiment, the NA61/SHINE
experiment is aimed to measure the pion and kaon yields for the T2K
experiment \cite{Antoniou:2006mh,Abgrall:2007zza}. Here the beam
energy is 30\GeV{}, while also upgrades to 40\GeV{} and 50\GeV{} are
foreseen. Responsible for the neutrino flux will be pions and kaons
with momenta $1\cdots10\GeVc$ and angles
$\theta\lesssim300\UNIT{mrad}$.

\begin{figure}[tb]
  \begin{center}
    \hspace*{\fill}%
    \includegraphics[width=7.5cm,clip=true]{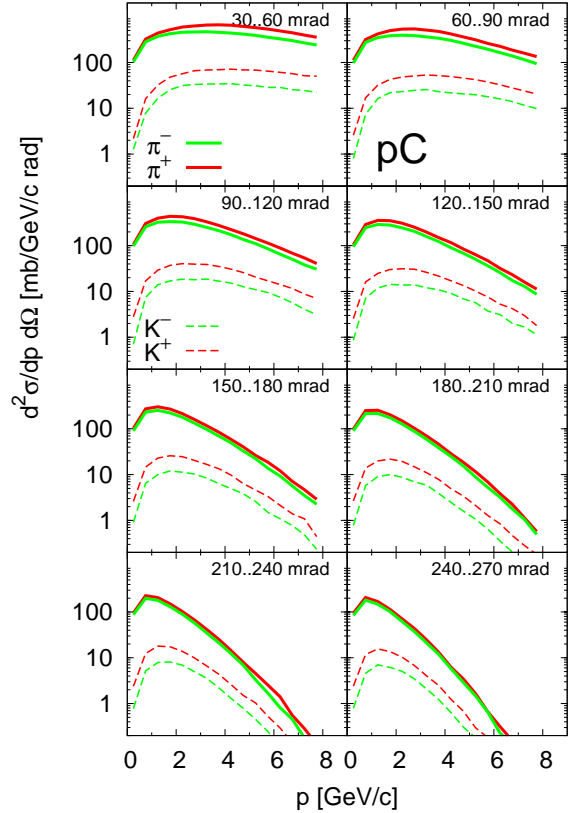}
    \hspace*{\fill}%

    \caption%
    {\textit{Cross section $d^2\sigma/dp\,d\Omega$ for
        $p+C\to\pi^\pm+X$ and $p+C\to K^\pm+X$ with 30\GeVc{} beam momentum
        as for the NA61/SHINE experiment.
        Thick lines show the yields of charged pions, while thin
        dashed curves indicate the yields of charged kaon states.
      }}
    \label{fig:Shine_C30}
  \end{center}
\end{figure}

In order to ease the comparison with the previous parts, we will use
the same binning for the calculations at these higher energies as
those for HARP with the forward spectrometer. In
fig.~\ref{fig:Shine_C30} we show our resulting
spectra of charged pions and kaons for a proton beam on a Carbon
target. We note that a comparison of the calculated pion yields with
24\GeVc{} on a proton target with experimental data
\cite{Blobel:1973jc} is as good as for 12\GeVc{} beam momentum
\cite{GiBUUGallery}. Therefore the uncertainties of the calculations
are again mainly due to the final state interactions. As soon as
experimental binning is available, we are able to produce the spectra
directly comparable with the upcoming data.

\section{Conclusions}

We have presented a comparison of calculations within the GiBUU
transport model with recent data on inclusive pion production on
different nuclear targets with pion or proton beams in the region of 3
up to 12\GeVc{} beam momentum. Contrary to other theoretical frameworks
we are able to
cover the full energy range of the HARP experiment.

The best description is achieved for the data with pion beams. The
agreement obtained for proton beams is very good over the whole
energy-range, except for very forward and very backward directions.
These deficiencies seem to be due to FSI as a comparison with
corresponding data for elementary $p + p$ collisions shows. This
underlines the need to understand results on elementary collisions
before drawing conclusion on data taken on nuclei as targets.

While the experimental data are especially useful as input to neutrino
flux calculations, we also have here a very powerful set of data for
checking final state interactions within our transport code, not only
as needed for the neutrino production processes, but also for
describing the interactions of neutrinos with nuclei in one and the
same theory and code.

We have also presented first theoretical results for the 30 GeV proton
run in the NA61/SHINE experiment which aims for a precise
determination of the neutrino flux in the T2K experiment. Since soon
the data from this experiment will cover a wide range of beam
energies, it will be interesting to extract from the beam energy
dependence conclusions on formation time aspects and color
transparency \cite{Gallmeister:inProgress}.

The authors would like to thank the whole GiBUU team for inspiring
discussions on pion production and final state interactions. We
gratefully acknowledge support by the Frankfurt Center for Scientific
Computing, where parts of the calculations were performed.



\begin{thebibliography}{99}

\bibitem{Catanesi:2008zz}
  M.~G. Catanesi et~al. (HARP Collaboration), Astropart. Phys. 29 (2008) 257.

\bibitem{Catanesi:2008zzb}
  M.~G. Catanesi et~al. (HARP Collaboration), Phys. Rev. C 77 (2008) 055207.

\bibitem{Catanesi:2005rc}
  M.~G. Catanesi et~al. (HARP Collaboration), Nucl. Phys. B732 (2006) 1.

\bibitem{GiBUU} http://theorie.physik.uni-giessen.de/GiBUU

\bibitem{Buss:2006yk}
  O.~Buss, L.~Alvarez-Ruso, A.~B. Larionov, and U.~Mosel, Phys. Rev. C74 (2006)
  044610.

\bibitem{Buss:2006vh}
  O.~Buss, L.~Alvarez-Ruso, P.~Muhlich, and U.~Mosel, Eur. Phys. J. A29 (2006)
  189.

\bibitem{Falter:2004uc}
  T.~Falter, W.~Cassing, K.~Gallmeister, and U.~Mosel, Phys. Rev. C70 (2004)
  054609.

\bibitem{Leitner:2006ww}
  T.~Leitner, L.~Alvarez-Ruso, and U.~Mosel, Phys. Rev. C 73 (2006) 065502.

\bibitem{Leitner:2008fg}
  T.~Leitner, O.~Buss, U.~Mosel, and L.~Alvarez-Ruso, PoS(Nufact08)
  (2008) 009.

\bibitem{Gallmeister:2005ad}
  K.~Gallmeister and T.~Falter, Phys. Lett. B630 (2005) 40.

\bibitem{Gallmeister:2007an}
  K.~Gallmeister and U.~Mosel, Nucl. Phys. A801 (2008) 68.

\bibitem{GiBUUGallery}
  http://gibuu.physik.uni-giessen.de/GiBUU/wiki/HarpGallery

\bibitem{Blobel:1973jc}
  V.~Blobel et~al. (Bonn-Hamburg-Munich Collaboration), Nucl. Phys. B69 (1974) 454.

\bibitem{Chemakin:2001qx}
  I.~Chemakin et~al., Phys. Rev. C 65 (2002) 024904.

\bibitem{Antoniou:2006mh}
  N.~Antoniou et~al. (NA49-future Collaboration) CERN-SPSC-2006-034.

\bibitem{Abgrall:2007zza}
  N.~Abgrall et~al. (NA61 Collaboration) CERN-SPSC-2007-033.

\bibitem{Gallmeister:inProgress} K.~Gallmeister and U.~Mosel, work in
  progress.

\end{thebibliography}
\end{document}